\documentclass{aastex}
\usepackage{emulateapj5,natbib,psfig}

\newcommand{\gtsim}{\mbox
{{\raisebox{-0.4ex}{$\stackrel{>}{{\scriptstyle\sim}}$}}}}

\slugcomment {accepted for publication, Astrophysical Journal}
\shorttitle{Rotation measure and source structure}
\shortauthors{}

\begin{document}

\title{Lowering Inferred Cluster Magnetic Field Strengths - the radio galaxy
contributions} 

\author{Lawrence Rudnick\altaffilmark{1} \& Katherine M.\
Blundell\altaffilmark{2}}

\altaffiltext{1}{University of Minnesota, Minneapolis, Minnesota, U.S.A.}
\altaffiltext{2}{University of Oxford, Astrophysics, Keble Road, Oxford,
OX1 3RH, U.K.}

\begin{abstract}
    
 We present a detailed examination of the relationship between the
 magnetic field structures and the variations in Faraday Rotation
 across PKS\,1246$-$410, a radio source in the Centaurus cluster of
 galaxies, using data from Taylor, Fabian and Allen.  We find a
 significant relationship between the intrinsic position angle of the
 polarization and the local amount of Faraday Rotation.  The most
 plausible explanation is that most or all of the rotation is local to
 the source.  We suggest that the rotations local to cluster radio galaxies may
 result from either thermal material mixed with the radio plasma,
 or from thin skins of warm, ionized gas in
 pressure balance with the observed galaxy or hot cluster atmospheres. 
  We find that the contribution of any unrelated
 cluster Rotation Measure variations on scales of 2 $-$ 10 arcsec are
 less than 25 rad/m$^{2}$; the standard, although model dependent,
 derivation of cluster fields would then lead to an upper limit of
 $\approx 0.4 \mu {\rm G}$ on these scales.  Inspection of the
 distributions of Rotation Measure, polarisation angle and total
 intensity in  3C\,75, 3C\,465 and Cygnus
 A also shows source-related Faraday effects in some locations. 
 Many effects  can mask the signatures of
 locally-dominated RMs, so the detection of even isolated correlations
 can be important, although difficult to quantify statistically.  In
 order to use radio sources such as shown here to derive {\it cluster-wide}
 magnetic fields, as is commonly done, one must first remove the local
 contributions; this is not possible at present.

\end{abstract}


\section{Introduction}
\label{sec:intro}

The Faraday Rotation of linearly polarized radiation through a
magnetic plasma has been used as a diagnostic of conditions in
synchrotron sources since the work by e.g., \citet {Bur66}.  When the
path length through the plasma and the density of thermal electrons
can be reasonably estimated, the observed Faraday Rotation, described
by the Rotation Measure (RM) in units of rad/m$^{2}$ can be used to
derive a characteristic magnetic field strength.  This technique has
been applied to clusters of galaxies, leading to estimates of
partially disordered cluster-wide fields which vary between less than
1 $\mu {\rm G}$ [e.g.\ \citet{Law82} and \citet{Hen89}] to 1$-$40$ \mu
{\rm G}$ [e.g.\ \citet{Kim91}, \citet{Ge93}, \citet{Fer99},
\citet{Cla01}, \citet{Tay01a}, \citet{Tay01b} \& \citet{All01a}].

In order to infer magnetic field strengths of intra-cluster media
(ICM) from the Faraday rotation of the polarised synchrotron radiation
from radio sources it is necessary that either a) the radio sources
are actually behind clusters and compared to a properly defined
control sample not seen through clusters, or b) if the radio source is
actually embedded within a cluster, that the contributions within or
caused by the source (``local'') are identified and isolated from any
diffuse cluster-wide Faraday effect.  These conditions have often been
violated in the existing literature.

In this paper we discuss how RM effects local to the source
 might be identified, with
particular reference to the case of a radio source in the Centaurus
cluster, PKS\,1246$-$410.  Our study includes in
Section~\ref{sec:rm_v_p0} a description of the signatures in the
RM-polarization angle (PA) plane for various origins of the RM
variations.  In Section~\ref{sec:simulatated_rm}
 we present a re-analysis of the polarimetric
data on this source published by \citet{Tay01a}.  We find that the RMs
are related to polarized features of the source itself, and therefore
locally caused, and not an indicator of an ICM field, as they
presumed.

In Section~\ref{sec:others}, we present a brief inspection of
relationships between RM and source structure in three well-known 3C
radio galaxies.  We discuss in Section~\ref{sec:implications} possible
physical origins for the local Faraday medium around cluster radio
galaxies.

\section{Signatures of source-related RM contributions}
\label{sec:rm_v_p0}

The derivation of cluster-wide magnetic field strengths, from 
observations of individual radio galaxies in the manner
of  \citet{Fer99}, 
\citet{Tay01a} \& \citet{Tay01b} relies on the assumption that the
radio sources' own Rotation Measures do not make a significant
contribution to the observed Rotation Measure from which the
cluster-wide field is derived.  

This assumption has been justified because total intensity structures
typically do not appear to correlate with RM variations.  However,
total intensity structures do not correlate well with polarized
emission, in general, because the latter is a vector quantity whose
variations in angle are significant sources of structure and
constructive and destructive interference.  A change in magnetic field
direction has the potential to change both the polarized emission and
the Rotation Measure, if sufficient thermal electrons are present. 
This is the basis of the search described here for source related RM
variations.

In the remainder of this section, we describe how a Rotation
Measure - polarization position angle (RM-PA) scatter diagram,
where each point samples a different position in the source,
can be used to diagnose the origin  of the RM variations in that
source.  In the next section, we will apply this analysis to
PKS\,1246$-$410.

\begin{figure}
\epsscale {0.5}
\plotone{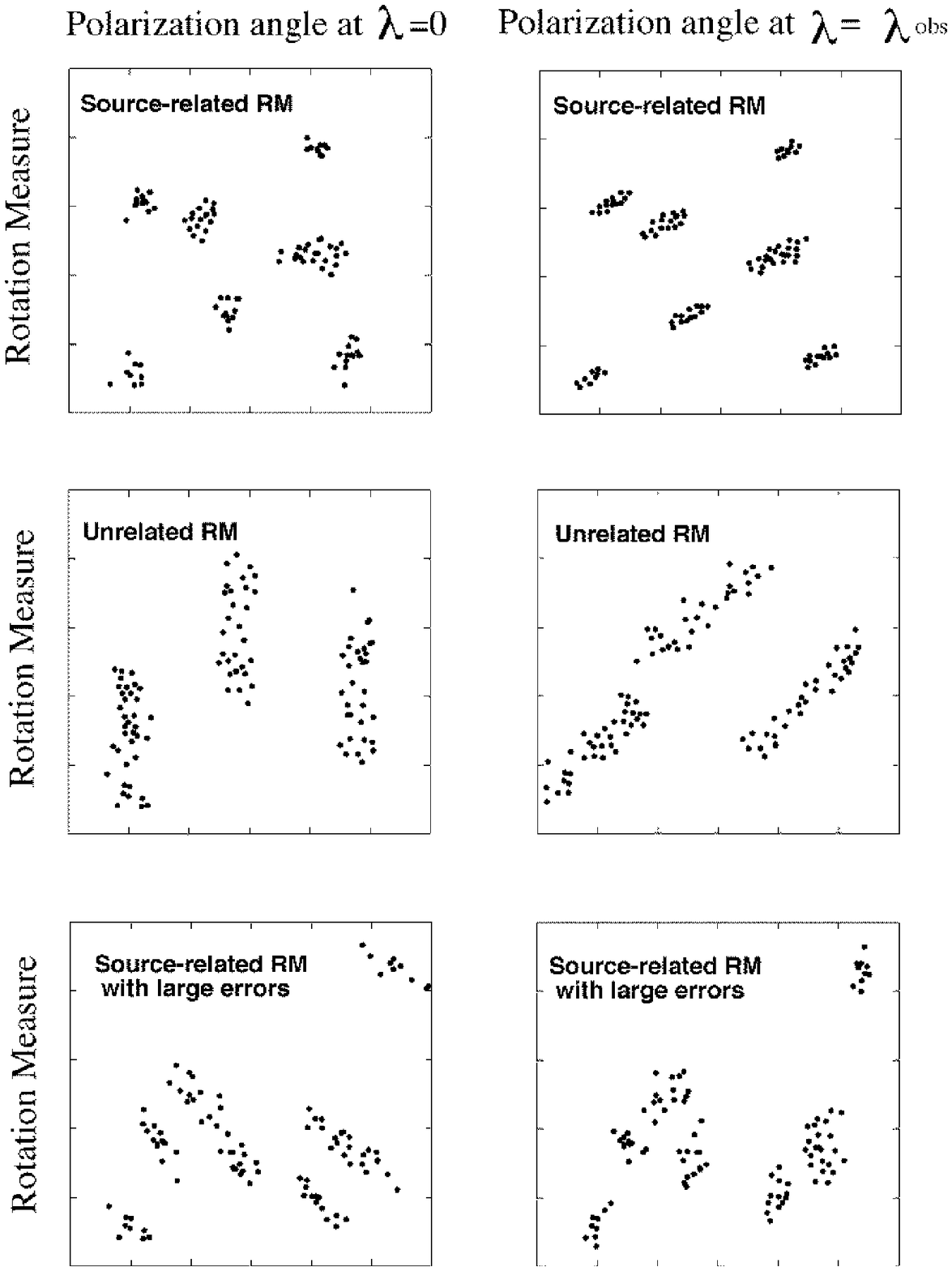}
\figcaption{\label{fig:cartoon} Indication of the expected dependences
of rotation measure and polarisation angle for different scenarios.}
\end{figure}

Figure\,\ref{fig:cartoon} explores the signatures in
the  RM-PA  plane that would arise
from various origins of any observed RM variations.  For each case, we
present two different plots: 1) RM-PA$(\lambda_0)$, which is the
scatter diagram of RM vs.\ the zero wavelength polarization position
angle {\it as calculated using the} RM, and 2) RM-PA$(\lambda_{\rm
obs})$, where the polarization angle is simply that observed at some
wavelength $\lambda_{\rm obs}$.

The top cartoons first assume that the radio source is composed of a
finite number of well-resolved ordered field regions, and that these
ordered fields are embedded in a Faraday rotating medium (which might
be internal to the source, or on a thin skin).  We then make the
simplifying and favorable  assumption  that
most observed magnetic field changes in the plane of the sky (giving
the PA) will be accompanied by some change along the line of sight
(giving the RM).  In such a situation, one would then see clusters of
points in the RM-PA$(\lambda_{0})$ diagram.  Note that in this
situation there is not a monotonic relation between PA and RM, {\it
i.e.,} there is no global correlation between the two quantities, only
a clumpy distribution.  

In the non-idealized case, the clumps will have a finite size for a
variety of reasons -- errors in the measurement of either quantity,
changes in the magnetic field strength along the line of sight that
are not well  correlated with PA changes, and changes in path
length or density of the Faraday rotating medium, which will change
the RM, but not PA$(\lambda_{0})$.  If any of these confusing factors
is sufficiently large, the Faraday rotation may still be physically
dominated by the source magnetic structure, but there will be a great
deal of scatter in the RM-PA diagram and signatures will be masked.
Looking for clumpiness in the PA-RM diagram is thus a one-sided test
-- if sufficient clumpiness is observed, the RMs originate local to
the source; if clumpiness is not observed, this test in isolation does
not provide information on the origin of the rotations. Some
clumpiness in the RM-PA plane can arise from source structure alone,
so this must be carefully evaluated, as discussed below.

Turning now to the correlation of RM with the PA observed at
$\lambda_{\rm obs}$, the clumps are stretched and repositioned by the
rotation at finite wavelength.  At sufficiently large values of
$\lambda_{\rm obs}$, the PAs will be scattered over all possible
angles.  As illustrated here, the clumps in the PA-RM plane can be
seen at both $\lambda_{0}$ and $\lambda_{\rm obs}$.

We now examine the signatures of RMs that arise externally and are
unrelated to the source.  If the RMs vary on small scales (a beam size
or less) and are completely unrelated to the source, one simply gets a
scatter diagram in the PA-RM plane.  This case is not shown. However,
the middle set of cartoons illustrates the important situation where
there are well-resolved regions of constant polarization angle in the
source, as well as well-resolved regions of RM, but which are now
completely unrelated to the source.  Because of the large scale
structures in both parameters, there will be some clumping in the
PA-RM plane.  To visualize this, suppose there were only three
intrinsic polarization angles observed in a source, each covering
one-third of the source.  Then the diagram would show three clumps in
the RM-PA$(\lambda_{0})$ diagram, stretched along the RM axis.  In the
RM-PA$(\lambda_{\rm obs}$) diagram, these clumps would be rotated and
displaced. It is important to separate this misleading clumping, due
to unrelated large scale structures in the source and in the external
RM distribution, from true correlations.  Such an effort is described
in the following section.

The final (bottom) set of cartoons in Figure\,\ref{fig:cartoon} shows
the effects of small errors in the RM measurements.  Errors within
each clump will tend to stretch the clumps in the PA$(\lambda_{0})$
plot to give negative slopes, because the wrong corrections have been
applied to the data.  Observations at $\lambda_{\rm obs}$ appear
similar to the no-error case (top cartoon), but with greater scatter
in the RM parameter.

In order to understand the role of measurement 
errors more fully, we ran a variety
of Monte Carlo simulations.  We examined the potentially
biased situation where coherent large scale errors in PA are actually
used to derive the RM, which is then plotted vs.  the same PA. In all
cases, the distributions in the PA-RM plane looked like either the
middle or bottom cartoon in Figure\,\ref{fig:cartoon}.  In no case
did errors or unrelated RMs cause small-scale clumping, i.e., clumps
with an extent significantly smaller than the overall range of RMs.
Although this gives us confidence that our results and
interpretation are reasonable, it does not guarantee robustness
against all possible distributions of PAs and RMs; this is discussed
further below.

\section{Analysis of PKS\,1246$-$410 and Results}
\label{sec:simulatated_rm}

We re-examined the polarimetric data on PKS\,1246$-$410 of
\citet{Tay01a}, kindly provided by G.\ Taylor, including their RM
image, which they derived using maps at 4 frequencies (8.4681, 8.1149,
4.8851 and 4.6351\,GHz).  We then created polarization angle at
$\lambda_0$ maps by de-rotating the position angle image at
8.4681\,GHz using Taylor et al's RM map.  \citet{Tay01a} blanked their
RM map wherever the residual polarization angle, $\delta {\rm
PA}(\lambda_0)$, at any frequency was greater than 15 degrees.  The
resolution of their images is 2.1\,arcsec by 1.2\,arcsec with a beam
position angle of 19 degrees.

In this section, we describe in detail the process we used to reach
the conclusion that the RM variations in PKS\,1246$-$410 had a
strong contribution local to the source.  First, we describe the
construction of simulated maps for comparison with the actual
data.  Then we use a variety of methods for comparison.  Finally,
we discuss the limitations to this analysis, and the inferences
for the Faraday rotating medium in PKS\,1246$-$410.

\subsection{Simulated map construction}
We created random simulated data to approximate as closely as possible
the amplitude and spatial distributions of the RMs measured by
\citet{Tay01a}, but which were not linked to source structure (in
total or polarised intensity) in any way.  These random images were
constructed by using the noise-generating feature of the AIPS task
IMMOD to create a large noise image, from which three independent
simulations could be drawn.

The first step was to reconstruct the amplitude distribution of RMs, 
which we did by adding together three large, independent  gaussian noise images each 
with a different mean RM and rms RM, 
to reflect the broad and narrow features in the
observed RM amplitude distribution shown in \cite{Tay01a}'s Figure 5.
The simulated RM distribution was then convolved spatially with a
circular gaussian of 5\,arcsec.  This value was picked by trial and
error, so that the spatial scales of the simulated RM distributions
matched that of the observed RM distribution.  A region of the large
noise map with the same number of pixels as the actual maps was
arbitrarily selected. Then, exactly as for
the observed RM map, the simulated random RM map was masked so that
only the structure within the silhouette of the source remained.  

\begin{figure}
\epsscale {1.0}
\plotone{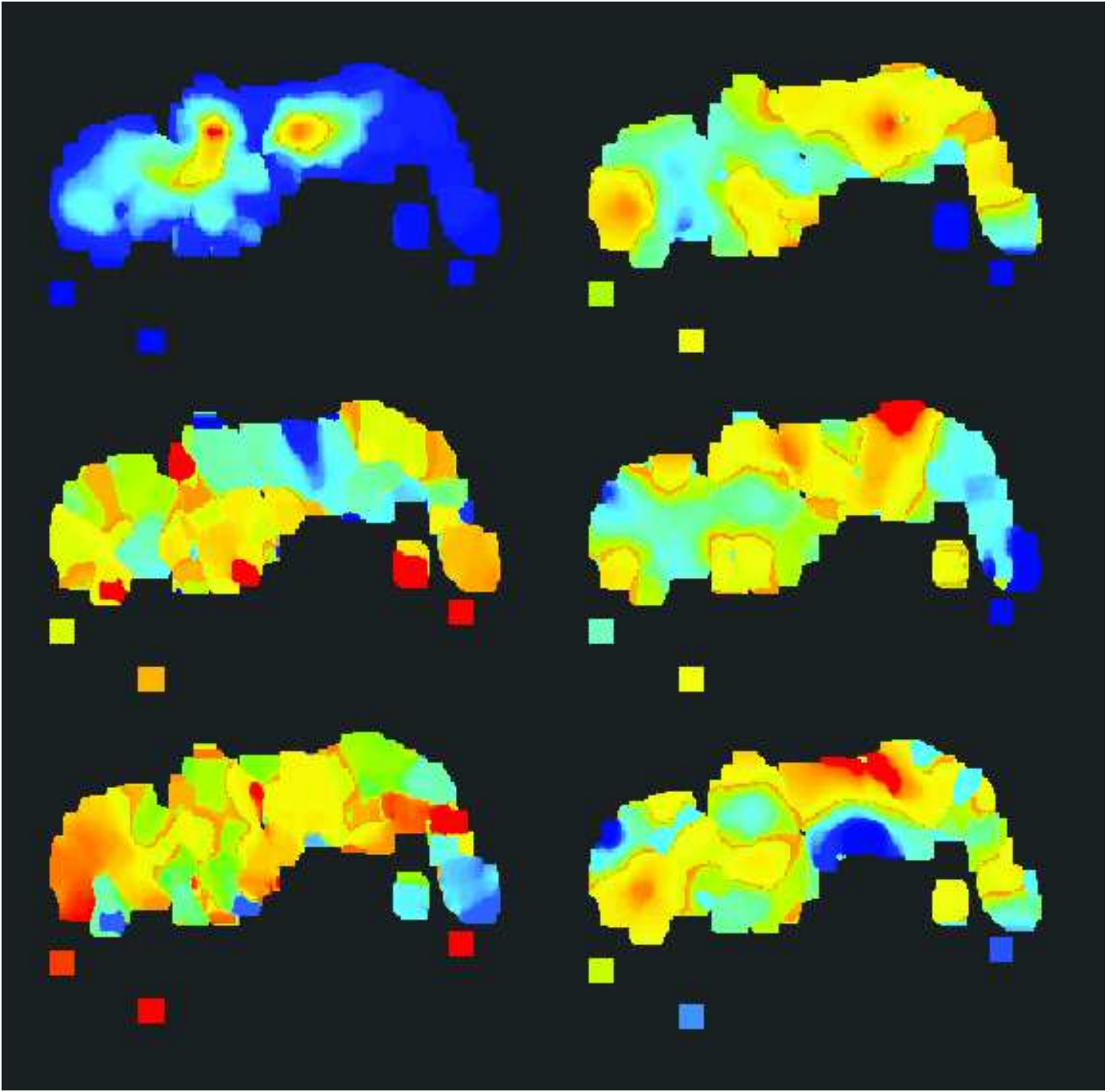}
\figcaption{\label{fig:colscale_gallery} Distribution of polarisation
structures in PKS\,1246$-$410. The source is $\approx 30\arcsec$ long;
North is directly up.  Left column $--$  actual data from Taylor et al.
(2001) top: polarized intensity, at 8 GHz, peak flux density 4 mJy/beam (red);
 middle: rotation measure, $\approx -1300$(blue) to $1300$(red) rad/m$^2$;
 bottom:
RM-corrected position angle ($-90$(blue) to $+90$(red) degrees. 
Right column $--$ simulations of rotation
measure, with same color scale as actual rotation measure at left middle.
  See Section 3.1 for processing information.    
}
\end{figure}

\begin{figure}

\plotone{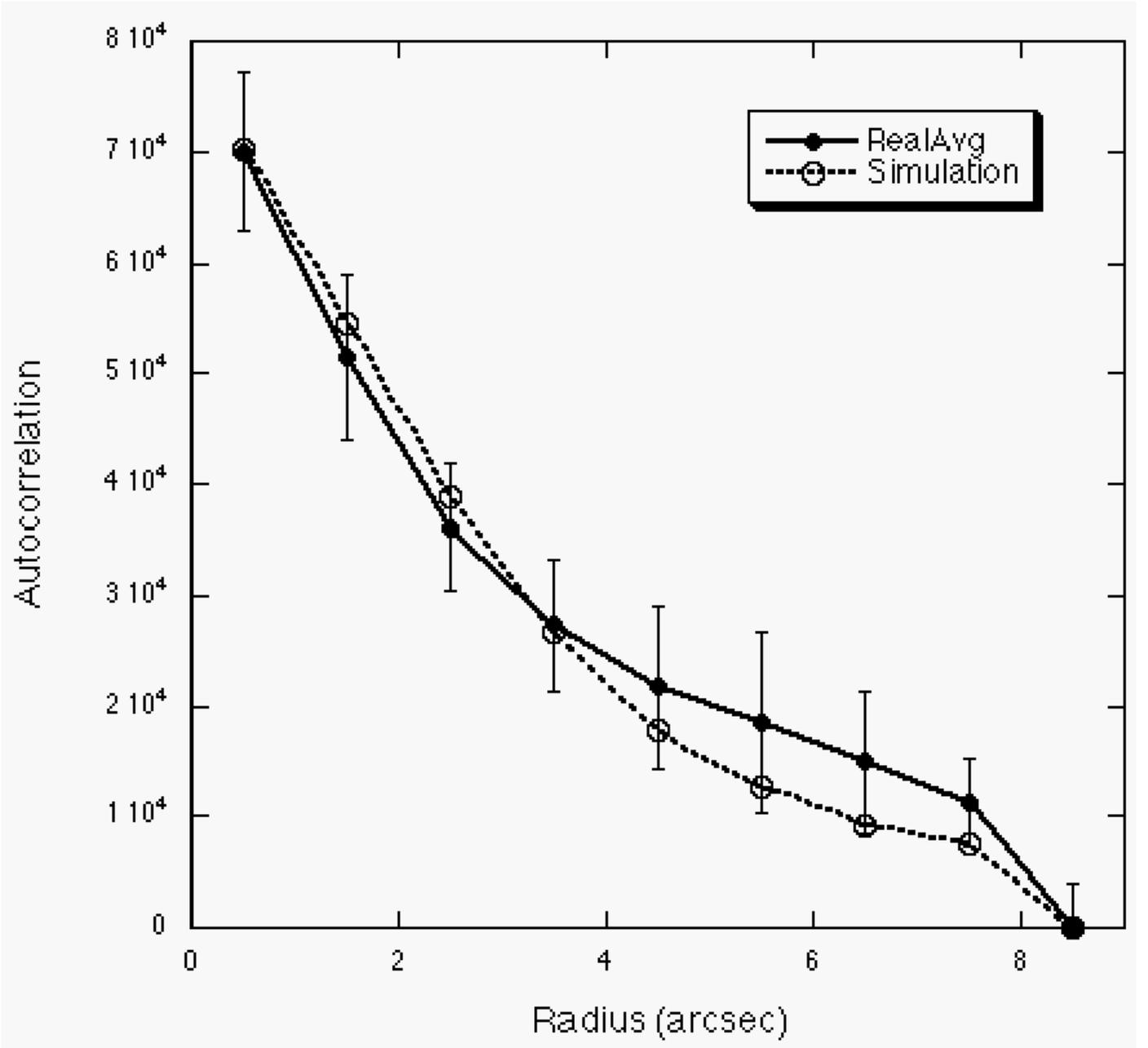}
\figcaption{\label{fig:simreliability} Autocorrelation functions of the
real and simulated RM distributions showing that they have the same
spatial scale structures. The simulated autocorrelation shown  is the
average of the three individual simulation correlations.}
\end{figure}

Three independent regions on the noise map were arbitrarily chosen
to create three such simulated RM maps.
 Finally, both simulated and observed maps
were median weight filtered using a box size of 3.2 arcsec (11
pixels), to reduce some of the high spatial frequency scatter.  The
similarity of the simulated and observed RM distributions can be seen
directly in Figures\,\ref{fig:colscale_gallery} and
\ref{fig:simreliability}.  Figure\,\ref{fig:colscale_gallery} shows
the similarity in amplitude and spatial scale of the observed and
simulated RM distributions.  Figure\,\ref{fig:simreliability} uses an
autocorrelation function to show quantitatively that the spatial
scales are well matched.

\subsection{Differences between observed and simulated RM maps}
\label{sec:real_v_sim}

In this section, we will examine the differences between the
observed and simulated maps in a variety of ways -- visual
inspection, a statistical analysis of the clumping in the RM-PA
plane (similar to two-point correlation analyses), and finally
a one-dimensional correlation/concentration analysis to determine
the typical scatter of RMs within clumps in the RM-PA plane.

\begin{figure}

\plotone{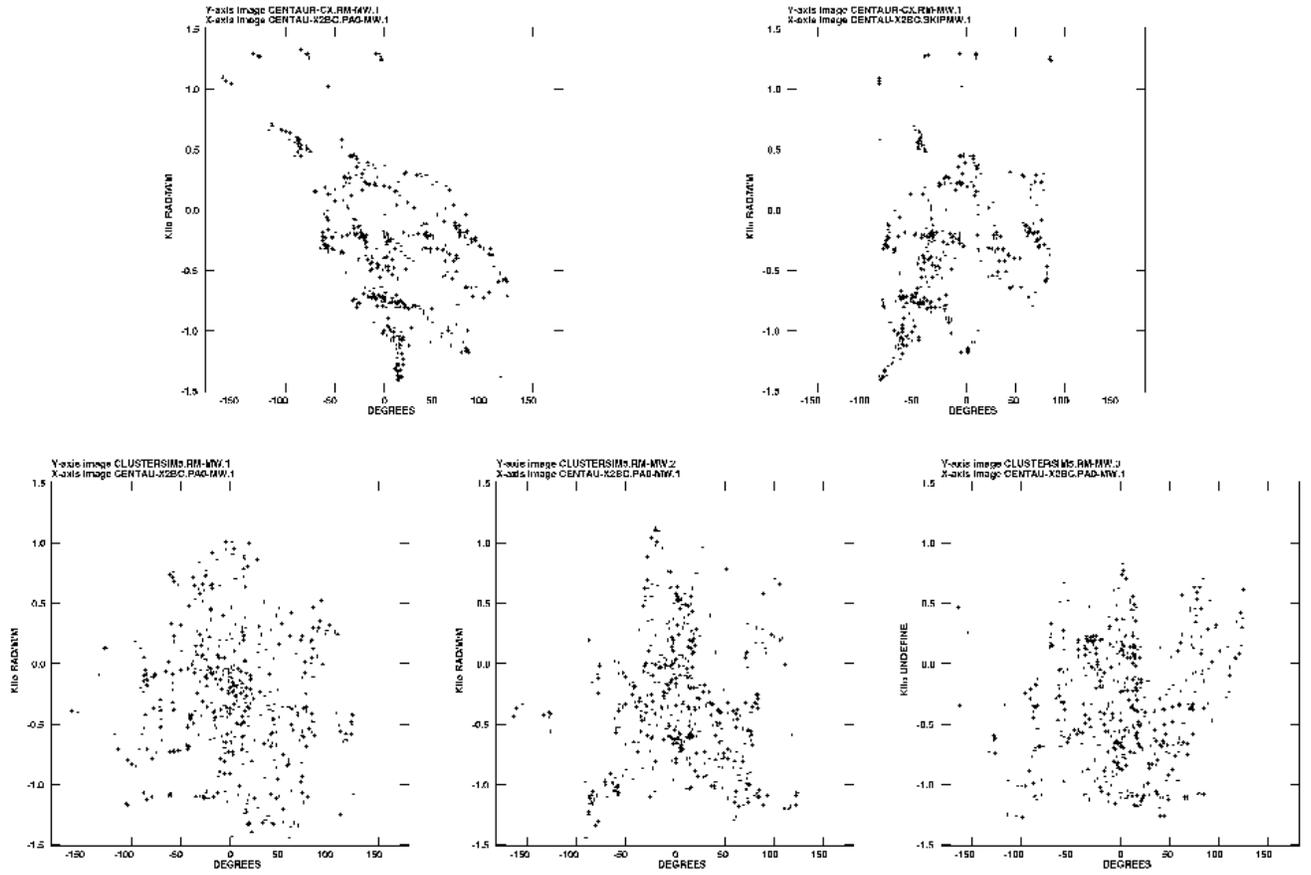}
\figcaption{\label{fig:scatters} Scatter plot of rotation measures
versus polarisation position angles.  Top left: actual data as observed
at 8.4\,GHz;  Top right: actual data at 8.4\,GHz with angles 
corrected to zero wavelength, using  RMs as plotted;  Bottom: 
simulations, with angles corrected to zero wavelength, as above.
 See Section 3.1  for description of  median-weighted filtering and processing. 
}
\end{figure}

Visual inspection of the middle and bottom maps on the left of
Figure\,\ref{fig:colscale_gallery} allows one to pick out
corresponding features in the actual PA and RM images.  At the same
time they indicate how nicely the RM simulations (on the right) mimic
the real data without showing any correlations with source structure. 
This can be seen more clearly in the scatter plots 
(Figure~\ref{fig:scatters})
of RM versus
polarisation angle, both corrected for Faraday rotation
[PA$(\lambda_0)$] and
uncorrected [PA$(\lambda_{\rm 3.6cm})$].
 For the RM-corrected position angles
(actual data) we also produced scatter plots versus each of the three
independent RM simulations; all plots are shown in
Figure~\ref{fig:scatters}.  For these plots, the data are somewhat
oversampled in order to ensure that the sharp observed transitions in
PA and RM are included; the actual data and simulations are sampled in
exactly the same way.

\begin{figure}

\plotone{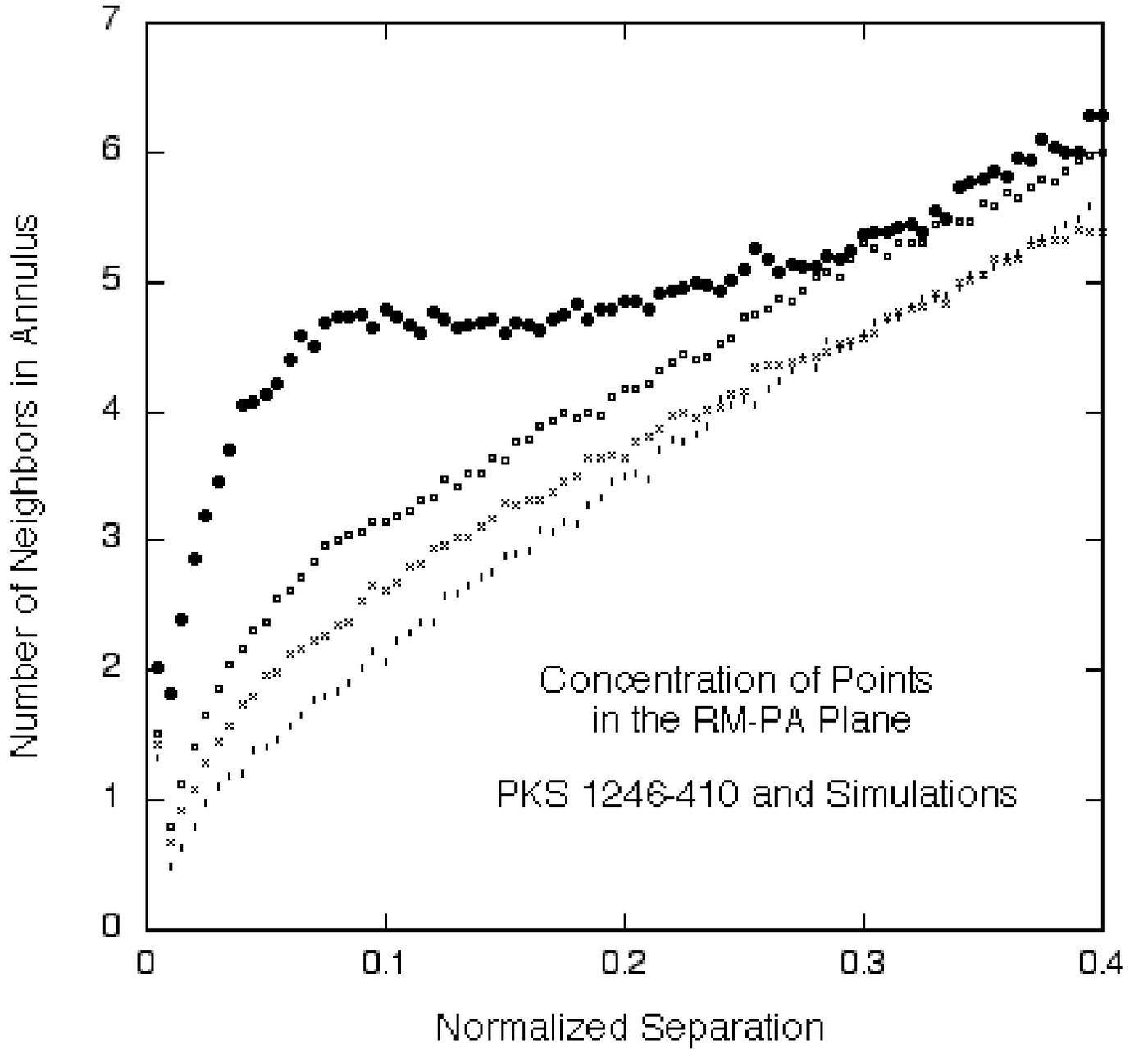}
\figcaption{\label{fig:concentration} Measurement of the clustering of
points in the  polarisation angle $-$ rotation measure plane 
(Figure\,\ref{fig:scatters}) by counting
neighbors in annuli as a function of normalized distance, as described
in Section 3.2.  The real data are shown as the large solid circles, while the
three individual simulations are shown with smaller symbols. }
\end{figure}

A visual inspection makes it apparent that the observed data are much
more highly clumped than the simulations.  In order to examine this
clumpiness more quantitatively, we calculated the average number of
neighbors around each point in the RM-PA plane, after normalizing each
of the PA and RM distributions to zero mean and unit variance.  A more
rigorous analysis would account for the $180$ degree ambiguity in PA,
but would not affect the current results. 
Figure\,\ref{fig:concentration} shows a large excess of neighbors for
the real data, compared to the simulations, up to a normalized
separation of at least 0.2.  The excess at the low end demonstrates
that the intrinsic source structure [as indicated by PA$(\lambda_0)$]
is related to the observed RM. A separation of 0.2 corresponds to a
difference in PA between two points of 45\,deg or 45\,rad/m$^2$ in RM,
or an equivalent combination of the two.  These values indicate the
scale of both the intrinsic scatter and the errors in RMs within each
fairly homogeneous region.

In the idealized limit of the source consisting of a finite number of
completely homogeneous components with random field strengths and
angles, partially mixed with a thermal plasma, we would expect to see
a finite set of tight clumps in the PA$(\lambda_0)$ versus RM diagram,
similar to what is observed.  If the RM structure were external to the
source, e.g.\ in the intracluster medium and not connected to the
source, then we would expect to see no such clustering in the
PA$(\lambda_0)$-RM plane other than the modest non-uniformities due to
the source structure itself, as illustrated in Figure
\ref{fig:cartoon}.  These results indicate that the RMs for
PKS\,1246$-$410 are related to its own polarization structure.

Note that the clumping is seen for the actual data whether or not
RM-corrections have been applied (i.e.\ for both PA$(\lambda_{\rm
3.6cm})$ and PA$(\lambda_0)$); therefore the derived RMs are not the
cause of the clumping.  However, in the RM-corrected position angles
we note that there are a variety of clumps which show elongation along
a slope of $-0.07$ to $-1.1$ degrees per rad/m$^{2}$ (at $\lambda
3.6$\,cm).  This is exactly as expected if small RM errors in each
major clump contributed to a spread in derived intrinsic position
angles, as seen in the bottom cartoon of Figure \ref{fig:cartoon}. 
The RMs are thus significantly more uniform in each region than
actually observed.  In addition, there is a slight tendency for the
overall distribution of clumps to be elongated in the same direction
as within individual clumps.  This could indicate errors in RM on
larger scales, although with only a small number of independent
clumps, it is not possible to establish this clearly.

\begin{figure}

\plotone{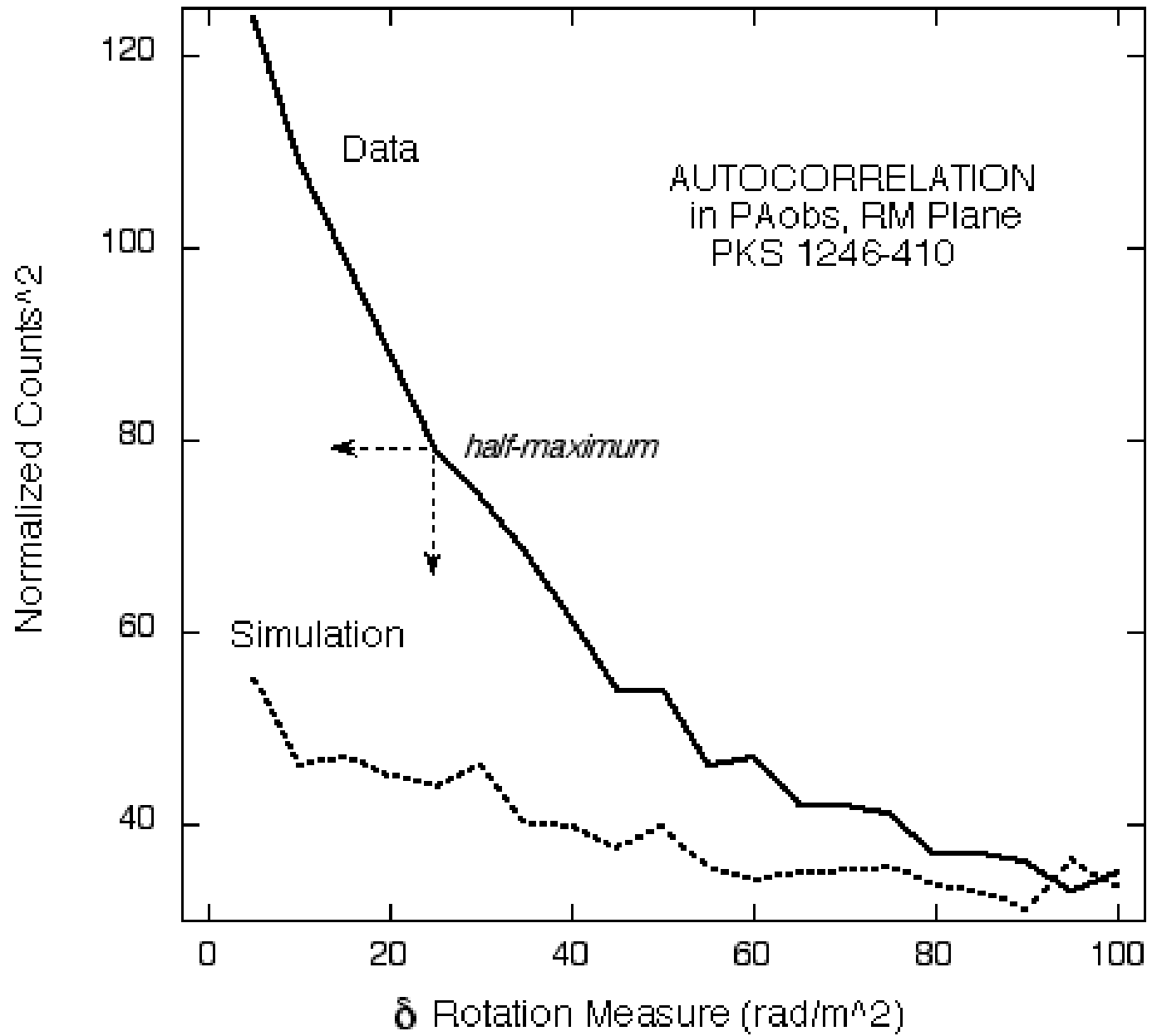}
\figcaption{\label{fig:autoRM} Autocorrelation as a function of shift in 
RM for the data and first (left) simulation plotted in 
Figure \,\ref{fig:scatters}. The dashed lines identify the approximate 
half-power point of the RM shift above the background autocorrelation
level.}
\end{figure}

We can use the clumping in the PA-RM plane to place a upper limit on
the contributions to the RMs in each region from external Faraday
media, unrelated to the source.  As discussed above, the width of each
clump along the RM axis in the PA-RM plane is dominated by small
errors, and we can therefore take this width as a conservative limit
to the contribution of external RMs.  To measure a characteristic
width of clumps in the entire plane, we form the one-dimensional
autocorrelation of the counts in PA-RM bins as a function of a shift
in the RM ($\delta RM$).  We first count the number of pixels in bins
of 1\,$\deg$ in $PA_{\rm obs}$ and $5\,{\rm rad/m}^{2}$ for both
PKS\,1246$-$410 and the first of our simulations, without any further
smoothing, and then sum their one-dimensional autocorrelations.
Note that this sum of one-dimensional
autocorrelations at each PA is not the same as the autocorrelation 
of the RM distribution summed over all PAs.
The results are shown in Figure \ref{fig:autoRM}.  The clumping in the
PA-RM plane is visible here as an excess at small values of $\delta
RM$.  At large values of $\delta RM$ the autocorrelation falls to the
same value as the simulation, reflecting the overall distribution of
RMs.  The half-width of the excess at small $\delta RM$s is $\approx
25\,{\rm rad/m}^{2}$, which we therefore take as an upper limit to
external contributions.

\subsection{Limits to the analysis}

Our derived external RM limit applies only over a limited range of
scale sizes.  Any external RM variations on scales $< 3 \arcsec$ would
be averaged out by the beam size, and so would not be detected in our
analysis.  At the other extreme, a constant external RM over the
entire source would simply move the clumps to a different location in
the RM $-$ PA plane, and again, not be detectable.  External RMs on
scales significantly larger than the coherent field regions ({\it
i.e.},$\gtsim$ 15\,arcsec) would thus not be identifiable.  The only
way to sample such large scale fields is by using the RMs of
well-constructed samples of background sources.

Although we have argued that the observed PA-RM clumping is easily 
distinguishable from the effects of random external RM distributions, 
this is based on a reasonable, but simple, set of assumptions.  In
particular, we assume that we can model uncorrelated external cluster
RM contributions by  random gaussian noise matched in both its
RM  amplitude distribution  and in its  distribution of RM patch
scale sizes.  We have shown by Monte Carlo simulations, again 
based on random gaussian fluctuations, that measurement errors do
not create the observed clumping.  We do not, however, prove that
our analysis is robust against all possible RM and PA distributions,
especially involving higher order correlations.  For example, we
have not modeled the possible effects of  large scale gradients in the RMs, or 
a tendency for extreme RMs to be closer together or further
apart than expected at 
random.  Such an open-ended exploration is beyond the scope of this 
paper.  It is also not clear what the utility of such an exploration
would be;  if one finally found a distribution that reproduced the
observed RM-PA clumping, it would simply show that an external RM
source is possible, not necessarily likely.  However, 
 it would be interesting to explore  physically motivated
models of cluster magnetic field structures, and determine their 
 manifestations in the PA-RM plane, and whether these could be distinguished from
local source effects.

\subsection{Inferences regarding PKS\,1246$-$410's  Faraday rotating medium}

Converting the RM limit of $\approx 25\,{\rm rad/m}^{2}$ into a limit
on the magnetic field in the cluster medium around PKS\,1246$-$410 is
a model-dependent process as discussed by \citet{New02}.  One must
make assumptions about the dependence of both the electron density and
the magnetic field strength as a function of the distance from the
cluster center, and the coherence length of the magnetic field.  As
shown by \cite{New02}, reliable measurements or limits to the field
depend on conducting Monte-Carlo simulations of the expected RM
distribution under a given set of assumptions about the cluster.  Such
a calculation is beyond the scope of this work, and we simply use our
upper limit in the identical way to \citet{Tay01a} for comparison with
other numbers in the literature.  \citet{Tay01a} use the overall RM
dispersion of $660\,{\rm rad/m}^{2}$ and X-ray measurements of the
Centaurus cluster from \citet{All01b} and \citet{San02} to estimate a
strength of $11 \mu {\rm G}$ for magnetic fields which vary on scales
of 1 kpc ($3.6 \arcsec$).  Using our limit of $25\,{\rm rad/m}^{2}$
for external contributions, we conclude that the strength of magnetic
field variations in the intracluster medium on this scale is $< 0.4
\mu {\rm G}$.  We note again that such a derivation is highly model
dependent.

\begin{figure}

\plotone{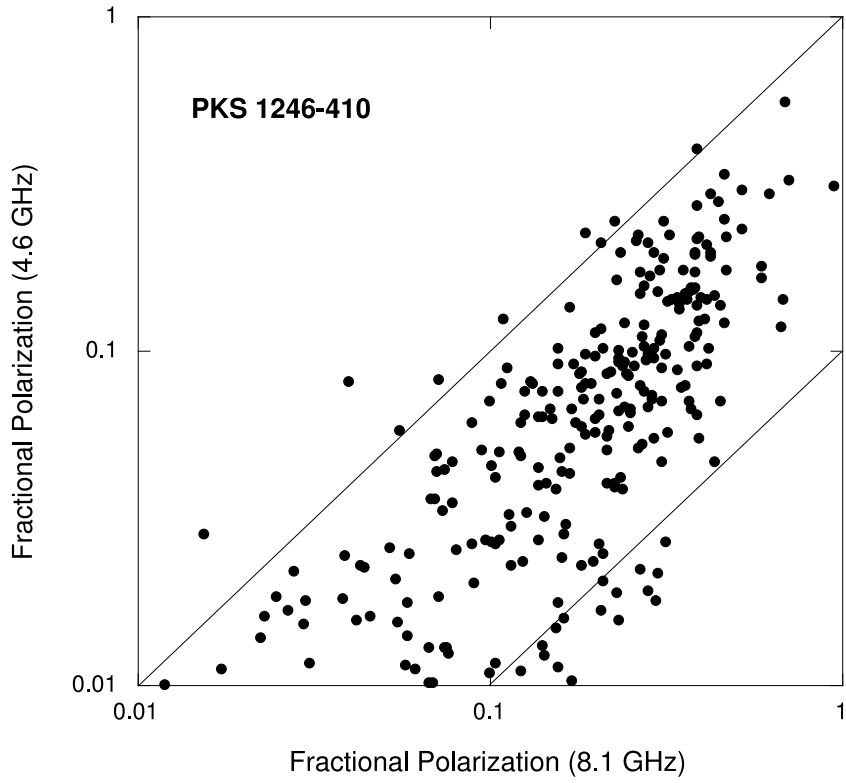}
\figcaption{\label{fig:depol} Fractional polarization at 4.6 GHz vs.
that at 8.1 GHz.  Lines of no depolarization (D=1) and depolarization
by a factor of 10 (D=0.1) are shown.  Note both the very high 
fractional polarizations present and the strong average depolarization. 
There are approximately 2 points plotted per independent beam area.  
Pixels that were blanked for the RM image, as described in Section 3.1, 
are not included here.}
\end{figure}

Further information about the Faraday rotating medium comes from
the observed depolarization between 8.1 GHz and 4.6 GHz.
 Figure\,\ref{fig:depol}
shows the fractional polarizations between these two bands plotted 
against each other.  Two things are worth noting -- the very high 
fractional polarization present in some parts of Taylor et al's image and the
strong mean depolarization ($\%Pol_{4.6GHz}/\%Pol_{8.1GHz} \approx 0.3$).

There are two general ways in which sources can become depolarized,
due to internal or ``external'' variations in RM \citep{Bur66}. 
We start with the  less popular alternative of internal depolarization
(e.g., \citet{Jag87}).  In this case, a thermal plasma
is mixed with the relativistic plasma, and along any line of sight the
rotation from regions near the back of the source will be rotated by
different angles from those near the front, with more destructive
interference  at longer wavelengths.  We can examine the
plausiblity of this idea by supposing that the X-ray emitting gas,
\citep{All01b,San02} with central density $n_{0} \approx 0.1
{\rm cm}^{-3}$ has penetrated throughout the source. The maximum
RM (and depolarization) will occur in the limiting, although
unrealistic, case of a uniform field from front to back, without
field reversals.
  Using a minimum energy field strength of $\approx 10
\mu {\rm G}$ and a line of sight path length of $\approx 3$\,kpc would
yield a total RM of 1400 ${\rm rad/m}^{2}$ through the source,
comparable to the largest values observed.  At 4.6 GHz, this
corresponds to a $\lambda^{2}$-equivalent rotation of $\approx 340
\deg$, which would correspond to a front-to-back depolarization by a
factor $\approx 40$ in the simplest slab geometry (e.g.,
\citet{Bur66,Cio80}). On the basis of these numbers alone, one could
try to construct more realistic geometries for an internal depolarization
model.  However, internal depolarization models result in very limited
rotations in observed polarization angle (e.g., $45 - 90 \deg$, depending on
geometry, \citet{Cio80}), not the $\approx 360 \deg$ observed for
PKS\,1246$-$410.

In the case that the depolarization is caused ``externally'', this
could either be in a thin skin along the boundaries of the source,
in the medium directly influenced by the source, or in the completely
unrelated foreground.  In all cases, where the rotation measure changes
rapidly from one beam to the next, the depolarization will be strongest,
as is observed. A typical change of 600 rad/m$^2$ between RM clumps 
results in an angle change of $\approx 45 (135) \deg$ at 8.1 (4.8) GHz, resulting in much more destructive interference at the lower frequency.
Unfortunately, there is no clear signature in the depolarization to
isolate the location of the ``external'' medium, and thus whether it
is the  magnetic field at the source boundaries,
 or an intracluster field, that is
responsible for the depolarization.

\section{Correlated RM-PA variations in  other  sources}
\label{sec:others}

PKS\,1246$-$410 presented a fortuitously powerful laboratory in which 
to search for source-related Faraday effects, because of its very 
high fractional polarization.  In such a case, there are probably few 
if any field reversals along the line of sight.  Thus, wherever the
Faraday rotation is taking place in the source, be it a thin skin, or 
an extended distribution of thermal plasma, the magnetic field along the 
line of sight should be simply related to the observed polarization 
angle on the sky.  In most radio sources, with lower fractional 
polarizations, the situation is much less clear.

In general, we expect to see only an occasional correspondence between
the RM and local magnetic field direction, even if the rotation is
completely local to the source.  In addition to probing different
components of the magnetic field, only the RM is sensitive to
fluctuations in thermal density, and the field near the surface will
typically differ from the line of sight averaged field in the plane of
the sky.  Therefore, even isolated regions of correspondence can
provide evidence for RM contributions local to the source.

We have examined in less detail the correspondence between observed RM
and the polarization position angle in three 3C  sources, and find some closely
related features in the two parameters, as described below.  This has
important consequences for the use of RM variations in cluster radio
galaxies to derive values for the (unrelated) magnetic field
variations in the intracluster medium.  Such a derivation can only be done after
the RM contributions local to the source are removed, a task that
would be extremely difficult to accomplish.  In the presence of
source-related RMs, it is still possible to set upper limits to
cluster fields, but not to determine their magnitude.

\subsection{3C\,75 and 3C\,465}

\begin{figure}

\plotone{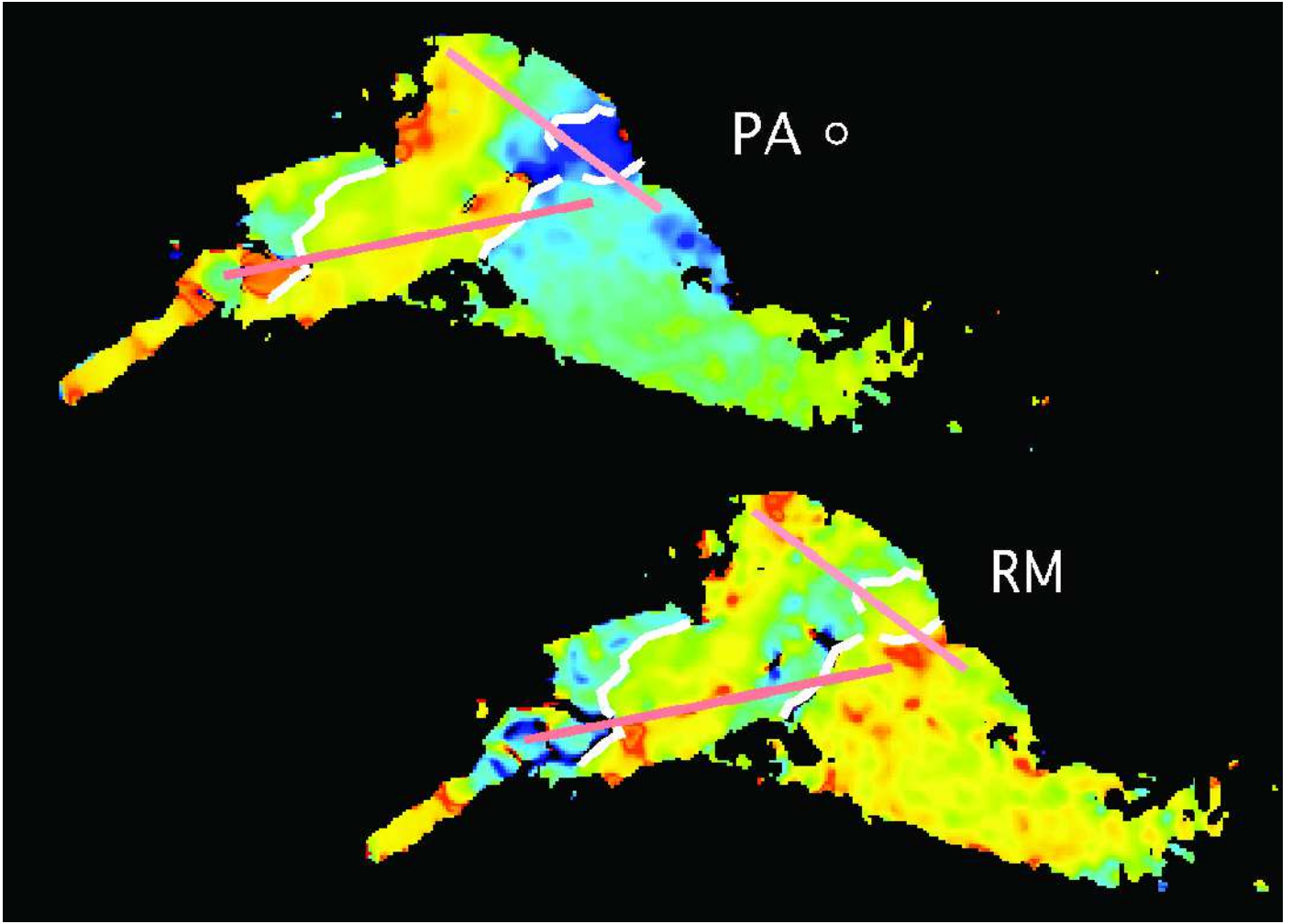}
\figcaption{\label{fig:3C465RM} Maps of the 
derived magnetic field direction (left,
upper) and RM (right, lower)      of the northern tail of
3C\,465 from \citet{Eil02}; the E-W extent of the source
is $\approx 120\arcsec$; North is directly up. Position angles vary from $-90$(blue) to $90$ (red)
 degrees.  Rotation measures vary from $\approx -200$ (blue) to $100$ (red)
rad/m$^2$.
White lines indicate some of the regions where there are 
corresponding transitions  in
both quantities. The off-white lines show the location of 
one-dimensional slices plotted in Figure \,\ref{fig:3C465slices}.
The beam size is indicated by the white circle.}
\end{figure}

Figure\,\ref{fig:3C465RM} shows the derived magnetic field direction
(left, upper) and RM (right, lower) maps of the northwestern tail of
3C\,465 from \citet{Eil02} and kindly provided by them.  The
northwestern tail provided a wide range of magnetic field directions,
spanning at least 120$\deg$ which we could examine for correlations
with the local RMs.  By contrast, the southern tail provided only
about 30$\deg$ of magnetic field angle variation, insufficient to
allow a correlation search.  However, \citet{Eil02} do note that even in
the southern tail, the extreme RMs ($\pm 200 {\rm rad/m}^{2}$) are found 
to be associated with the hot spot.

\begin{figure}
\epsscale {0.7}
\plotone{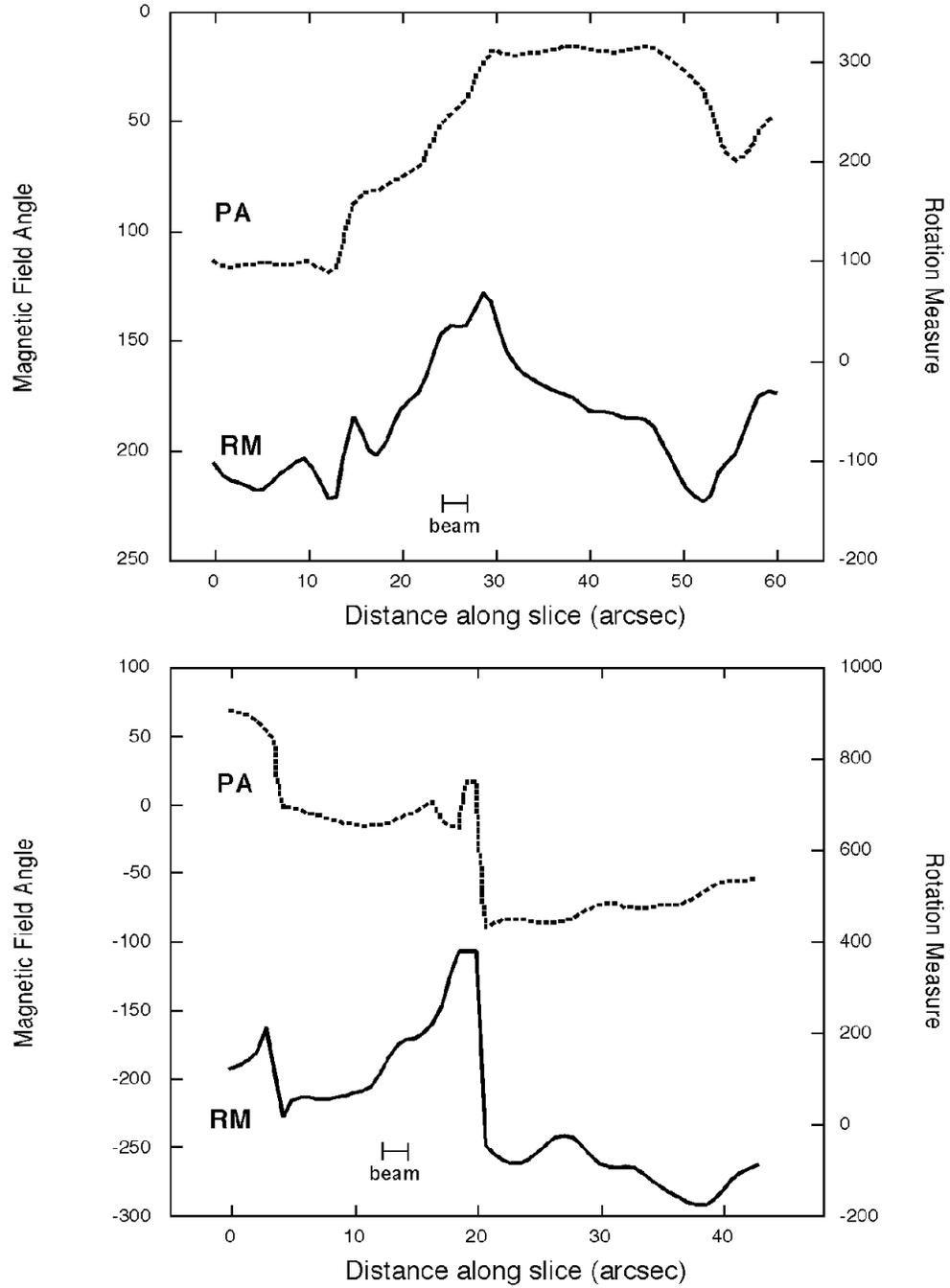}
\figcaption{\label{fig:3C465slices} One-dimensional slices in RM
and magnetic field direction along the lines shown in
 Figure 8 }.
\end{figure}

In the northwestern tail
we find some places where the magnetic field direction and the RM both
undergo sharp changes, as indicated in Figure\,\ref{fig:3C465RM}. 
These transitions can be more clearly seen in the one-dimensional
slices (Figure\,\ref{fig:3C465slices}) at the locations indicated in
Figure\,\ref{fig:3C465RM}.  Note that the beam size is much smaller
than the regions of coherent magnetic field direction.  The indicated
transitions in RM and magnetic field direction occur against a much
smoother background, and are not the result of random small-scale
variations.  These figures and slices demonstrate that a relationship
exists between the magnetic field direction and the RM at least in
some locations in 3C465, and therefore a significant part of the observed
Faraday rotating medium is local to the source, as in PKS\,1246$-$410.

\begin{figure}
\epsscale{1.0}
\plotone{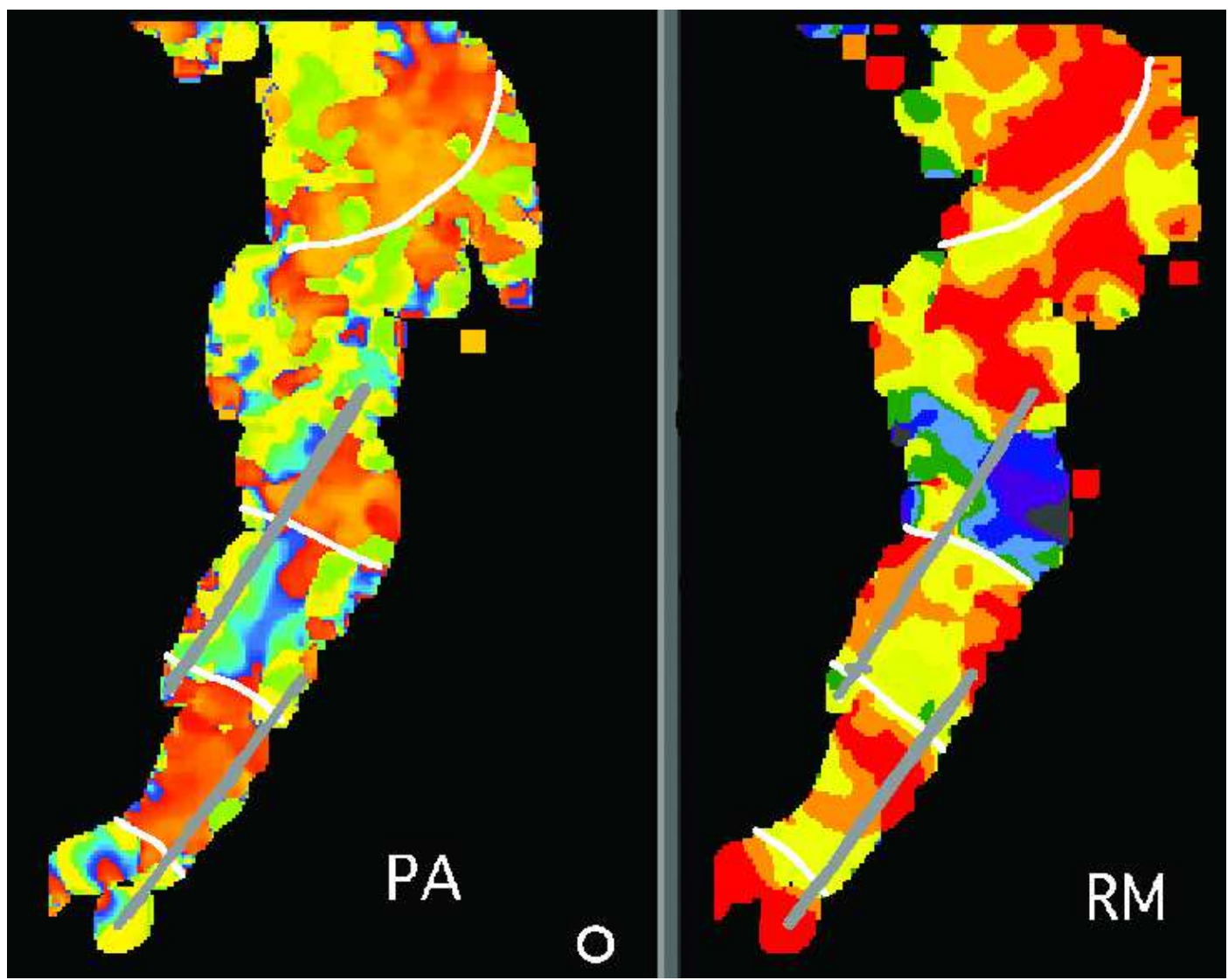}
 \figcaption{\label{fig:3C75RM} Same as Figure
\,\ref{fig:3C465slices}, but for the northern tail(s) of 3C75. 
 The N-S  extent of the source
is $\approx 110\arcsec$; North is directly up.
 Position angles vary from $-90$(blue) to $90$ (red)
 degrees.  Rotation measures vary from $\approx -100$ (blue) to $100$ (red)
rad/m$^2$. }
\end{figure}

\begin{figure}
\epsscale {0.6}
\plotone{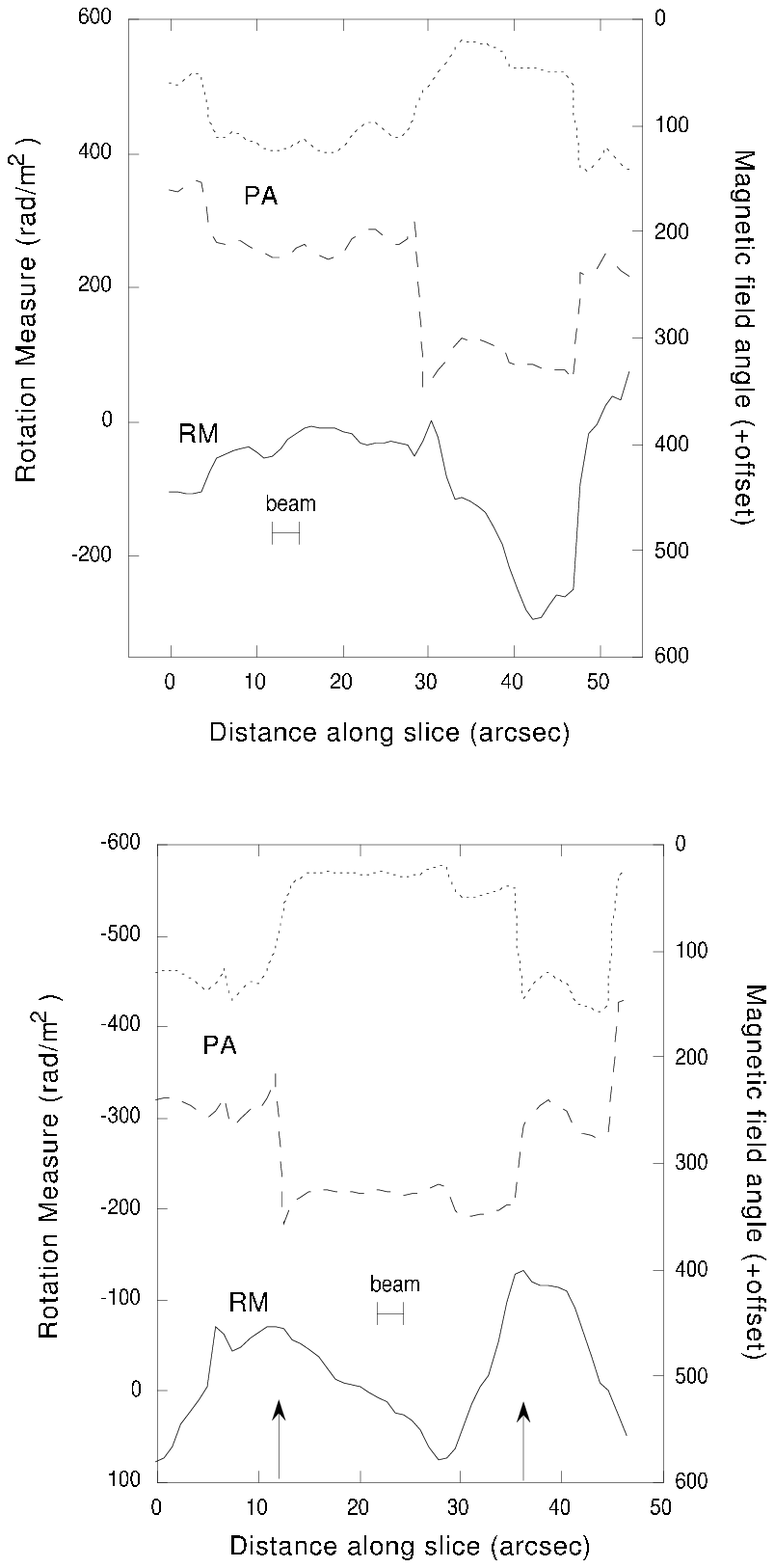}
\figcaption{\label{fig:3C75slices}  One-dimensional slices in RM
and magnetic field direction along the lines shown in
 Figure 10.  Two versions of the magnetic field 
 slices are shown, with different choices for resolving 180\,degree
 ambiguities.}
\end{figure}

3C\,75, again from \citet{Eil02}, provides another example of
correlated changes.  Only the northern tail(s) are shown here, because
they contained sharper transitions in magnetic field direction and
therefore clearer examples of correlated changes.  We note again that
given the various factors, as discussed in Section \ref{sec:rm_v_p0},
 that can independently influence the RM and
magnetic field angles, we expect only isolated correlations.  We first
smoothed the data with a median weight filter of five pixels (less
than one beam) which reduces some of the smallest scale mottling.  A
complication that arises in this analysis comes from the 180\,degree
ambiguities in magnetic field direction.  A smooth transition, e.g.,
from 170\,degrees to 190\,degrees will appear as a large jump if the
latter is recorded as $-10$\,degrees.  In our analysis of 3C\,75, we
found a number of apparent jumps (also see \citet{Eil02}) and
arbitrarily added 180\,degrees to smooth the magnetic field angles as
much as possible.  The resulting PA and RM distributions for the
northern half of 3C\,75 are shown in Figure\,\ref{fig:3C75RM}, with
the corresponding slices in Figure\,\ref{fig:3C75slices}.  There still
remains an overall 180\, degree ambiguity in the regions where the
magnetic field variations were smoothed.  Therefore, in the slices, we
show both possibilities, and note that the corresponding transitions
in magnetic field direction and RM appear in either case.

\subsection{Cygnus A}

\citet{Bic90a,Bic90b} find that at least some of the RM features in
Cygnus A are associated with features in the total intensity image
which they call surface waves. This is an unusual case, because total
intensity structures often show little or no correspondence to RMs.
 \citet{Per95} also find one specific RM - total intensity
relationship in the distinct  semi-circular region surrounding
 hotspot ``B'', which 
\citet{Car88} interpreted as due to a bow shock caused by the hotspot; 
\citet{Per95} do not find other RM - total intensity 
correspondences in the jet and hotspot
features.  

In the case of Cygnus A, \citet{Bic90a,Bic90b} propose that the
intensity related RM structures arise from non-linear Kelvin-Helmholtz
waves in a mixed magnetic field and thermal plasma.  These give rise
to large rotation measure variations on the scale-size of the waves.
Such a model would be consistent with the fact that the overall
pattern in RM is the same in each lobe.  Moreover, being a surface
effect, this would give rise to a dependence of RM on $\lambda^2$.
\citet{Dre87} observed such a  $\lambda^2$ dependence of polarization
angle, and eliminated {\em internal} Faraday rotation in Cygnus\,A;
they also eliminate a Galactic origin.  Although Dreher et al.'s
abstract states that they propose the origin of the rotation measure
is in the intra-cluster gas, their conclusions describe how the
Faraday screen could {\it equally well} be the intra-cluster gas {\em
or} a sheath around the lobes, a possibility we consider further, and
favour, in Section\,\ref{sec:implications}.

\section{Implications of a source-related Faraday medium}
\label{sec:implications}

We have shown that for PKS\,1246$-$410 the RM variations across the
source are likely to be 
dominated by a Faraday medium associated with the source
itself, with little or no cluster field contribution.  Indications of
source-related RM variations were also shown for 3C75, 3C465 and
Cygnus A.  These results have important implications both for studies
of intracluster magnetic fields as well as for the physics of radio
galaxies. In this section, we look briefly at issues such as the role of mass entrainment in
cluster sources, the presence of warm, dense optically emitting gas,
the trend for higher RMs toward cluster centers, the radio luminosity
dependence of depolarization, the side-to-side depolarization
asymmetries seen in some samples of radio galaxies and quasars, 
and statistical RM probes of cluster fields.

\subsection {Entrainment}

Entrainment is likely to be a common feature of FR\,I jets.  For
example, \citet{Bic94} has used the conservation laws to demonstrate
the relationship between entrainment and deceleration.  In a similar
manner, \citet{Lai02} find that mass-loading or entrainment of ambient
matter is required to reproduce the observed deceleration in the jets
of FR\,I radio galaxy 3C\,31.  Once thermal matter is entrained within
the magnetic field of the synchrotron emitting plasma it must at least
be considered as contributing to observed Faraday Rotation.  Without
considering entrainment, which could appear as correlations in the
RM-PA plane as discussed here, cluster magnetic field strength
estimates can get very high, for example, $\approx 10 \mu{\rm G}$ in
Abell 119 \citep{Fer99} or $\approx 35 \mu{\rm G}$ in Hydra A
\citep{Tay93}.  The current work shows that such derivations are not
justified.

In addition to direct mixing of the thermal and relativistic plasmas, 
radio sources in clusters can have a dramatic influence on the surrounding 
medium.  Holes in the hot X-ray emitting plasma are seen, for 
example, around radio sources in the Perseus \citep{Boh93,  Fab02} and Hydra
\citep{McN00} clusters.  Bouyant bubbles from Centaurus A also 
appear to have influenced the external medium \citep{Sax01}. 
Therefore, in addition to Faraday rotation from material which is 
directly mixed with the radio plasma, there could also be 
contributions from the ICM which are correlated with the radio source 
properties.  In such a situation, the magnetic field strengths derived 
from the RMs would apply to the ICM influenced by the source, but 
could not be used to measure cluster-wide fields.

\subsection {RM trends toward cluster centers}
There appears to be a trend that radio sources closer to the cluster
center show higher RM variations than those further out
\citep{Fer99,Tay01b,Gov01,Dol01}.  Although this has been interpreted as
demonstrating the existence of cluster-wide fields, two factors
indicate that a Faraday medium local to the source, as suggested here,
is also consistent with the data.  First, cluster sources are almost
exclusively FR\,Is, showing distorted morphologies that are usually
understood as due to mass loading and entrainment of the surrounding
medium (as discussed in Section 5.1).  Second, the environments are indeed
denser towards the centers of clusters, so even if the penetration of
source fields into a thin skin of thermal material were the same for
all sources, we would see larger Faraday effects in the vicinity of
the cluster centre.  Similarly, the high densities in cooling flow
clusters would lead to high rotation measures for any embedded sources,
(e.g., \cite{Tay01a}), independent of any cluster-wide fields.
It is critical to remove the dependence of
source-related RMs  on their local surrounding density before making any
claim for stronger fields towards the centers of clusters.

\subsection {Optically emitting gas as a Faraday medium} Several
theoretical arguments have been made to show that the RMs of embedded
cluster sources cannot plausibly arise local to the source.  One 
argument is that a fully
mixed thermal plasma would greatly depolarize the sources
\citep{Dre87}.  
 It has been suggested that
even a thin skin of thermal material
 can be problematic as the origin for RMs \citep{Tay93,Ge93,Eil02} , based either on geometric
arguments or the very high inferred strengths for magnetic fields near
the source boundaries.   However, most of these arguments are based on
the restrictive and unjustified assumption that the only thermal
material available is the diffuse hot X-ray emitting plasma.  Even in 
this restrictive case, Table 1 in \citet{Eil02} shows that a partially 
mixed hot plasma could produce the observed RMs in 3C75;  this is not 
reflected in their discussions or conclusion.  

Wherever one could document the implausibility of hot gas as the
Faraday medium, it is still necessary to consider the effects of gas
at $10^{4} - 10^{5}$K. There is an extensive literature documenting
the presence of warm, dense emission-line material in the centers of
rich X-ray clusters \citep{Hec81,Cow83,Hu85,Koe99}. 
 These emission-line systems are often found to be
interacting with radio galaxies on scales of 10s to over 100 kpc
\citep{Hec84,Bau89,McC90}, and in the cases of A
\,2597 and Coma \,A, for example, form a clear boundary layer around the
radio lobes \citep{Koe99,Tad00}.  Such material causes
regions of strong depolarization \citep{Hec84,Ped89a,Cha90} and
must therefore also contribute to the RM wherever the medium is
slightly more Faraday transparent.  

In situations where this gas is slightly less dense, or not yet
sufficiently cooled, it will not be prominent in
emission lines, but can still be a very effective Faraday medium with
a modest amount of field penetration into the radio source.  In 3C75,
for example, \citet{Owe90} use long-slit spectroscopy to set limits on
the density and temperature of cooler gas in pressure equilibrium with
the surrounding hot cluster medium.  For a 2.5\,kpc cloud, they find
that the temperature must be above $2.5\times 10^{5}$K, yielding a
density at pressure matching of $\approx 0.2 {\rm cm}^{-3}$.  If 3C75's
magnetic field were also in pressure equilibrium with this gas (6 $\mu
{\rm G}$ , \citet{Eil02}), then it would produce an RM of $\approx
2500 {\rm rad/m}^{2}$, 25 times higher than the RMs observed in this source.
This shows that sufficient warm material to produce the observed RMs is
still allowed by the observations;  it does not demonstrate that this
material actually exists.  We can also ask whether sufficient gas to 
cause the RMs could be at the more plausible temperature of $10^4$ K.
Using the same size cloud as above, and relaxing the pressure
balance criterion,  we need material only $0.04$ times
as dense as above to produce the observed RMs.  The emissivity would then go up
by $\approx 25$ because of the lower temperatures (see Fig. 7 in \citet{Owe90}), and down by $\approx 0.002$
because of the lower density, and would thus not have been detected.  
\citet{Owe90} also report observations of 3C465.  However, their slit did not cover any of the northern tail 
discussed in this paper.

Given the common appearance of emission line systems, and even the
currently available limits, we consider it quite plausible that the
cluster sources showing large RM variations are interacting with a
dense, warm medium.  Searches for additional emission line systems,
(e.g., \citet{Tad00}, \citet{Owe90}) and kinematics of this material
(e.g. \citet{Hu85}) could be of great benefit both in understanding
the physics of the radio sources and their energy input to the
intracluster medium.  Further exploration of sources showing
Faraday effects intrinsic or local to the source \citep{
Ped89b,Liu91,Bes98,Ish99} would also be useful, as well as demonstrating
why they cannot be used to derive unrelated cluster field strengths.

\subsection{RM trends in samples of radio galaxies}

\citet{Gar91}, in their Figure 2, show that the Faraday dispersion for
a sample of classical double FR\,IIs is an increasing function of $z$
(and thus luminosity in a flux-limited sample).  \citet{Pen00} show
how the fraction of high RM sources strongly increases with redshift.
Are these telling us about the large-scale environment of these
sources \citep{Car97,Pen00}, or about the properties of the sources
themselves?  Before a claim for environmental effects is possible,
corrections for luminosity effects would have to be determined and
made.  In all current models for radio-galaxy evolution, higher
luminosity sources have higher magnetic field strengths; in the
thin-skin Faraday rotation proposed here, this would automatically
lead to higher depolarizations and RMs in high-luminosity sources as
observed, even if the external densities were the same for all
sources.  It is therefore premature to interpret the
luminosity/Faraday relations as due to a denser environment at large
redshifts (luminosities), although that might be demonstrated to be
the case in the future.  A further claim that such relations imply a
large scale magnetized cluster gas, separate from the sources
themselves \citep{Car97} is an untested assumption until the
individual radio sources can be examined for source-related Faraday
contributions, as done in this paper.

Similarly, the lower level of depolarization seen in the (presumably
nearer, Doppler-boosted) jetted side compared to the non-jetted side
of some samples of radio sources (originally suggested by
\citet{Lai88} and \citet{Gar88}) has been cited as evidence for an
extended, magnetized halo (e.g., \citet{Gar91}).  However, we note
that such an asymmetry merely demonstrates (and requires) the
existence of a medium on the same scale as a particular radio source
that shows this effect,
and, in any case, requires magnetic fields of only $\approx 1 \mu{\rm
G}$ in a hot medium for sources with $1 \leq z \leq 2$. 

\subsection{Statistical searches for cluster fields}

In addition to attempts to measure Faraday rotation through the
intracluster medium using individual sources, as discussed above,
there have been a number of efforts to look for an excess RM using
background sources in the directions of clusters of galaxies compared
to samples away from clusters.  There are several sets of work
reporting incompatible results (e.g., \citet{Cla01}; \citet{Kim91};
\citet{Hen89}).  In a separate paper, we  present a thorough
analysis of these studies.  At present, we note that the current
paper's results on individual radio galaxies in clusters show that
some or all of their RM variations are local to the source, and can
thus not be used to measure cluster fields.  Unfortunately, the most
recent and extensive work on sources seen ``through'' clusters
\citep{Cla01} has a cluster sample that is dominated by radio galaxies
actually embedded in the clusters under study, i.e., not background
sources.  When all problematic sources are removed from the samples of
\citet{Hen89} and \cite{Cla01}, there is little, if any, evidence
remaining for cluster fields at the claimed $>1 \mu{\rm G} (> 0.1 {\rm
nT}) $ levels \citep{Rud03}.  \citet{New02} have provided an
additional critique of the statistics presented by these studies,
demonstrating serious problems with how RM limits or values are
translated into field strength estimates.

\section{Conclusions}
\label{sec:conc}

Given the existence of a Faraday medium connected to individual radio
galaxies, we conclude that the claims for strong intracluster magnetic
fields, 1 $-$ 40 $\mu{\rm G}$ ( 0.1$-$4 nT), based on studies of
individual radio galaxies are not warranted by careful
consideration of the current data.  The current best estimates
for cluster field strengths are in the $< 1 \mu{\rm G}$ ($< 0.1$ nT)
range, from cluster synchrotron halos or inverse Compton measurements
\citep[see refs in][]{Car02} and from theoretical arguments from the
early universe \citep{Bar97}.  There is, in addition, a great
opportunity to learn about the physics of radio galaxies from a more
detailed study of their Faraday rotating skins.

\acknowledgments

K.M.B.\ thanks the Royal Society for a University Research Fellowship. 
Partial support for extragalactic research at the University of
Minnesota is provided by the U.S. National Science Foundation under
grant AST 00-71167.  We thank Greg Taylor, Jean Eilek and Frazer
Owen for kindly allowing us to use their reduced images for these
investigations, and  Chris Carilli, Tracy Clarke, John Dickey, Jean Eilek,
Robert Laing, Frazer Owen and Greg Taylor for useful discussions. We
thank
an anonymous referee for helpful comments on the manuscript.  The VLA is a
facility of the NRAO operated by Associated Universities, Inc., under
co-operative agreement with the National Science Foundation.  This itself
research has made use of the NASA/IPAC Extragalactic Database, which
is operated by the Jet Propulsion Laboratory, Caltech, under contract
with the National Aeronautics and Space Administration.

\end{document}